\documentclass[aps,prd,amsmath,amssymb,twocolumn,preprintnumbers]{revtex4}
\pdfoutput=1
\usepackage{graphicx}
\usepackage{color}
\newcommand{\beq}{\begin{eqnarray}}
\newcommand{\eeq}{\end{eqnarray}}

\newcommand{\bmp}{\noindent\begin{minipage}{16cm}}
\newcommand{\emp}{\end{minipage}\vskip 7mm} 

\usepackage{dcolumn}
\usepackage{bm}
\usepackage{bbm}
\usepackage{subfigure}
\usepackage{pxfonts}

\definecolor{rossoCP3}{cmyk}{0,.88,.77,.40}

\def\lsim{\mathrel{\rlap{\lower4pt\hbox{\hskip1pt$\sim$}}
    \raise1pt\hbox{$<$}}}                
\def\gsim{\mathrel{\rlap{\lower4pt\hbox{\hskip1pt$\sim$}}
    \raise1pt\hbox{$>$}}}                

\newcommand{\be}{\begin{eqnarray}}
\newcommand{\ee}{\end{eqnarray}}

\preprint{CP3-ORIGINS-2010-XX}

\usepackage{fancyhdr}
\newcommand{\drawsquare}[2]{\hbox{%
\rule{#2pt}{#1pt}\hskip-#2pt
\rule{#1pt}{#2pt}\hskip-#1pt
\rule[#1pt]{#1pt}{#2pt}}\rule[#1pt]{#2pt}{#2pt}\hskip-#2pt
\rule{#2pt}{#1pt}}

\newcommand{\Yfund}{\raisebox{-.5pt}{\drawsquare{6.5}{0.4}}}
\newcommand{\Ysymm}{\raisebox{-.5pt}{\drawsquare{6.5}{0.4}}\hskip-0.4pt%
        \raisebox{-.5pt}{\drawsquare{6.5}{0.4}}}
\newcommand{\Ythrees}{\raisebox{-.5pt}{\drawsquare{6.5}{0.4}}\hskip-0.4pt%
          \raisebox{-.5pt}{\drawsquare{6.5}{0.4}}\hskip-0.4pt%
          \raisebox{-.5pt}{\drawsquare{6.5}{0.4}}}
\newcommand{\Yasymm}{\raisebox{-3.5pt}{\drawsquare{6.5}{0.4}}\hskip-6.9pt%
        \raisebox{3pt}{\drawsquare{6.5}{0.4}}}
\newcommand{\Ythreea}{\raisebox{-3.5pt}{\drawsquare{6.5}{0.4}}\hskip-6.9pt%
        \raisebox{3pt}{\drawsquare{6.5}{0.4}}\hskip-6.9pt
        \raisebox{9.5pt}{\drawsquare{6.5}{0.4}}}

\newcommand{\Yadjoint}{\raisebox{-3.5pt}{\drawsquare{6.5}{0.4}}\hskip-6.9pt%
        \raisebox{3pt}{\drawsquare{6.5}{0.4}}\hskip-0.4pt
        \raisebox{3pt}{\drawsquare{6.5}{0.4}}}
\begin{document}
\title{\Large  \color{rossoCP3} Flavor Dependence of the S-parameter}
\author{Stefano {\sc Di Chiara}$^{\color{rossoCP3}{\varheartsuit}}$}
\email{dichiara@cp3.sdu.dk} 
\author{Claudio {\sc Pica}$^{\color{rossoCP3}{\varheartsuit}}$}
\email{pica@cp3.sdu.dk} 
\author{Francesco {\sc Sannino}$^{\color{rossoCP3}{\varheartsuit}}$}
\email{sannino@cp3.sdu.dk} 
\affiliation{
$^{\color{rossoCP3}{\varheartsuit}}${ CP}$^{ \bf 3}${-Origins}, 
Campusvej 55, DK-5230 Odense M, Denmark.}
\begin{abstract}
We extend the results of \cite{Sannino:2010ca} by computing the S-parameter at two loops in the perturbative region of the conformal window. Consistently using the expression for the location of the infra-red fixed point at the two-loop order we express the S-parameter in terms of the number of flavors, colors and matter representation. We show that S, normalized to the number of flavors,  increases as we decrease the number of flavors. Our findings support the conjecture presented in \cite{Sannino:2010ca} according to which the normalized value of  
the  S-parameter at the upper end of the conformal window constitutes the lower bound across the entire phase diagram for the given underlying asymptotically free gauge theory. We also show that the non-trivial dependence on the number of flavors merges naturally with the non-perturbative estimate of the S-parameter close to the lower end of the conformal window obtained using gauge duality \cite{Sannino:2010fh}. Our results  are natural benchmarks for lattice computations of the S-parameter for vector-like gauge theories.  \\[.1cm]
{\footnotesize  \it Preprint: CP$^3$-Origins-2010-32}
\end{abstract}

\maketitle
\thispagestyle{fancy}

\section{Introduction}
Non-Abelian gauge theories are expected to exist in a number of different phases which can be classified according to the force measured between two static sources.
The knowledge of this phase diagram is relevant 
for the construction of extensions of the 
Standard Model (SM) that invoke dynamical electroweak (EW) symmetry
breaking \cite{Weinberg:1979bn,Susskind:1978ms}.   An up-to-date review is \cite{Sannino:2009za} while earlier reviews are \cite{Hill:2002ap,Yamawaki:1996vr}.  The phase diagram is also useful in providing ultraviolet completions of 
unparticle \cite{Georgi:2007ek} models \cite{Sannino:2008nv,Sannino:2008ha} and it has been investigated recently using different analytical methods \cite{Sannino:2004qp,Dietrich:2006cm,Ryttov:2007sr,Ryttov:2007cx,Sannino:2009aw,Gies:2005as,Poppitz:2009uq,Antipin:2009wr,Jarvinen:2009fe,Fukano:2010yv,Frandsen:2010ej}.

Here we wish to understand, in a rigorous way, the dynamics of gauge theories possessing an infrared fixed point. We therefore add a relevant operator, i.e. the fermion mass term to the theory assumed to possess large distance conformality. This is a standard procedure when trying to determine the properties of a generic fixed point. As discovered in \cite{Sannino:2010ca} the left-right spin one two-point function turns out to be an excellent probe of such dynamics thanks to the fact of being well behaved in the ultraviolet and in the infrared.

The language of the EW precision parameters is borrowed to connect more easily to the phenomenological world. In the first part of this work, which concerns perturbative results, we will not address the breaking of the EW symmetry and hence we choose the reference of the Higgs mass in such a way that the sole contributions to the precision parameters come from the calculable new sector. In other words we are computing, for the theory in isolation, the difference between the two- point functions of the vector-vector and axial-axial gauge bosons. 

In the second part of the Letter we go beyond perturbation theory and review the novel approach proposed in \cite{Sannino:2010fh} which makes use of electric-magnetic gauge duality.  Combining the perturbative and nonperturbative results we provide new evidence for the validity of the conjecture made in \cite{Sannino:2010ca} stating that the opportunely normalized $S$-parameter decreases with the number of flavors reaching the lowest value at the upper end of the conformal window. The stronger form of the conjecture predicts that the normalized $S$-parameter decreases as function of the number of flavors across the entire phase diagram. This conjecture can be falsified via lattice simulations within and outside the conformal window. The weaker form of the conjecture simply implies that the normalized $S$-parameter is bounded from below by its value at the upper end of the conformal window. 

Our results shed light on the dynamics within the conformal window and serve as important guide to numerical simulations of gauge theories displaying large distance conformality.

\section{Reviewing the Conformal S-parameter}
In \cite{Sannino:2010ca} one of the authors derived the one loop value of the $S$-parameter at the upper end of the conformal window where the perturbative expansion in the gauge coupling is reliable. 
We reiterate, to avoid possible misunderstanding, that the quantity which was studied in \cite{Sannino:2010ca} and we are interested in is the contribution to the vacuum polarizations coming solely from a new conformal sector in the presence of a mass deformation.

The oblique parameters $S$, $T$ and $U$~\cite{Peskin:1990zt,Peskin:1991sw,Kennedy:1990ib,Altarelli:1990zd} provide a sensitive test of new physics affecting the EW breaking sector. In this work we concentrate on the $S$-parameter, but it is straightforward to generalize the present analysis to all the other relevant parameters. The definition of $S$ we use is the same as in \cite{He:2001tp} which was also used in \cite{Sannino:2010ca}:
\be 
S&=&-16\pi\frac{\Pi_{3Y}(q^2)-\Pi_{3Y}(0)}{q^2} \,,
\label{eq:Sdefinition}
\ee
where $\Pi_{3Y}$ is the vacuum polarization of the third component of the isospin into the hypercharge current and we use as reference point, instead of the usual $Z$ boson mass, the external momentum $q$ of the gauge boson. 

We summarize the results of \cite{Sannino:2010ca} which made use of the 1-loop expression for $S$ to obtain a perturbative result at the upper end of the conformal window.

We consider  a sufficiently large number of flavors $N_f$ for which the underlying gauge theory develops an infra-red fixed point (IRFP) at a vanishingly small value of the coupling constant. In this regime the theory is perturbative as shown by Banks and Zaks  in \cite{Banks:1981nn}.

The quantum global symmetries are $SU_L(N_f)\times SU_R(N_f)\times U_V(1)$ if the fermion representation is complex or $SU(2N_f)$ if real or pseudoreal. To make contact with the SM, we assume $N_D=N_f/2$ doublets to be weakly gauged. Gauge and topological anomalies can always be canceled, if present, by adding new fermion doublets neutral with respect to the new dynamics.

At 1-loop the $S$-parameter is given by~\cite{He:2001tp}:
\begin{multline}
S= \frac{\sharp}{6\pi}\left\{2(4Y+3)x_1+2(-4Y+3)x_2-2Y\log\left(\frac{x_1}{x_2}\right)+\right.\\
+\left.\left[\left(\frac{3}{2}+2Y\right)x_1+Y\right] G(x_1)  
+\left[\left(\frac{3}{2}-2Y\right)x_2-Y\right] G(x_2) \right\}\,, \label{eq:Sfermion}
\end{multline}
with 
\begin{equation}
G(x)=-4\sqrt{4x-1}\,\arctan\frac{1}{\sqrt{4x-1}}
\,,
\label{eq:Gfun}
\end{equation}
 where in the above expressions $Y$ is the hypercharge, $x_i=(M_i/q)^2$, $i=1,2$, with $M_i$ the masses of up- and down-type fermions and  $\sharp = N_D \, d[r] $ is the number of doublets $N_D$ times the dimension of the representation $d[r]$ under which the fermions transform. 

Using the 1-loop expression of the $S$-parameter two independent and opposite limits can be taken:
in the first one we take the external momentum $q$ goes to zero keeping the fermion masses fixed; in the second one the fermion masses vanish at fixed $q$. These two limits do not commute as shown in \cite{Sannino:2010ca}. 

\subsection{ Sending $q^2$ to zero at fixed fermion masses}

In this limit, which is the relevant one for models of EW symmetry breaking, it was found in~\cite{Sannino:2010ca} that the $S$-parameter does {\it not} vanish inside the conformal window.

Taking $M_1=M_2 = m$, we obtain \cite{Sannino:2010ca}: 
\begin{eqnarray}
\lim_{\frac{q^2}{m^2}\rightarrow 0}S =  \frac{\sharp}{6\pi}\left[1 + \frac{1}{10x} + \frac{1}{70 x^2} + {\cal O}(x^{-3})\right] \ , 
\label{smallq}
\end{eqnarray}
with $ x=\frac{m^2}{q^2}$. 
Note that the leading term in the above formula for the $S$-parameter does not depend on the value of the fermion masses.
Moreover the dependence on the hypercharge $Y$ vanishes for $M_1=M_2=m$.


\begin{figure}[t]
\begin{center}
\includegraphics[width=\columnwidth]{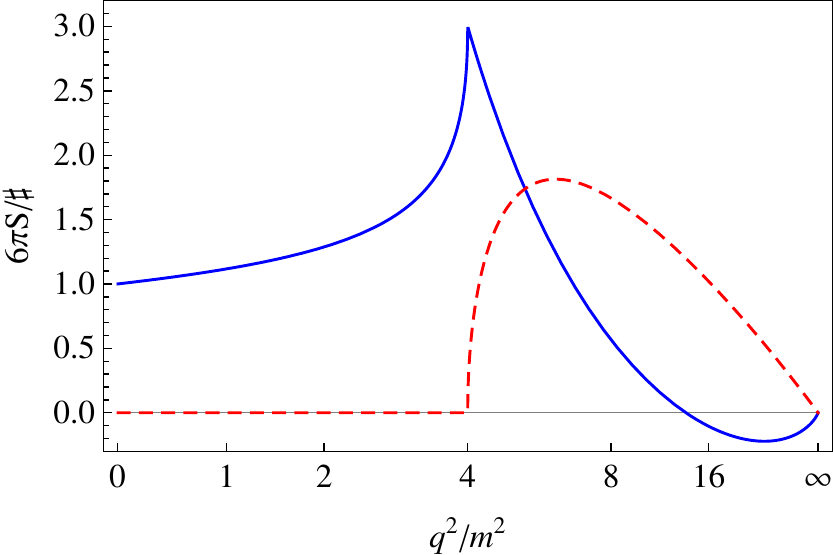} 
\caption{Real (blue, solid) and imaginary (red, dashed) parts for the normalized $\displaystyle{\frac{6\pi S}{\sharp}}$ parameter as function of increasing $q^2/m^2$ and $\sharp = \frac{N_f}{2}\, d[r] $. To plot simultaneously the $q^2/m^2 \rightarrow 0$ and $\infty$ limits  we use a nonlinear scale for the horizontal axis which is proportional to $\arctan \left( q^2/m^2 \right ) $. } 
\label{Sqlimit}
\end{center}
\end{figure}%
The reason why the $S$-parameter does not vanish in this limit is that the conformal limit is not reached when keeping the fermion masses fixed. This will in fact only be achieved in the opposite limit when we first send to zero the fermion mass while keeping the momentum finite (see below).
 
In Fig.~\ref{Sqlimit} we plot the complete 1-loop expression for the real (blue, solid) and imaginary (red, dashed) parts of the normalized $S$-parameter defined as $\displaystyle{{6\pi S}/{\sharp}}$.
Note that at the kinematic threshold $q^2 = 4m^2$ an imaginary part develops, which is associated to particle production in the fermion loop since the external momentum is sufficiently large to create, on shell, a fermion-antifermion pair.

\subsection{Sending $m^2$ to zero first and the conformal limit}

In the opposite limit $m^2/q^2\rightarrow 0$ one finds for the real and imaginary parts \cite{Sannino:2010ca}: 
\begin{eqnarray}
\lim_{\frac{m^2}{q^2}\rightarrow 0} \Re[S] & = &  x\,\frac{\sharp}{\pi}\left[2 + \log (x) \right] + {\cal O}(x^{2}) \ , \label{real}  \\
\lim_{\frac{m^2}{q^2}\rightarrow 0} \Im[S] & = &  x\,\sharp + {\cal O}(x^{2})   \  .
\label{ima}
\end{eqnarray}
Both $\Re[S]$ and $\Im[S] $ are nonzero but in this case they vanish with the mass when keeping fixed the external reference momentum $q^2$. 
This limit corresponds in Fig.~\ref{Sqlimit} to the $q^2/m^2 \rightarrow \infty$ region of the plot.
Note that due to the logarithmic term the $\Re[S]$ becomes negative before approaching zero. 

\section{Conformal $S$-parameter at 2-loops}

The 2-loops contribution to the $S$-parameter is given by:
\be
\Delta S=\frac{\alpha}{4\pi}\frac{\sharp}{6\pi}C_2\left[r\right] \delta S \,\, ,\label{EQS2LOOP}
\ee
where $\alpha$ is the coupling constant of the new sector, and $C_2\left[r\right]$ is the quadratic Casimir of the fermion representation. For completeness we give the full expression for $\delta S$ in the Appendix~\ref{APPA} corresponding to the 2-loops technicolor contribution to the $S$-parameter. This value is obtained by specializing to the case of degenerate fermion masses. This expression has been derived by starting from the results given by Djouadi and Gambino of the QCD corrections to the EW gauge boson self-energies~\cite{Djouadi:1993ss}. The main point here is the consistent application of these results to the conformal window which was not done in the literature before. In the main text we concentrate on the asymptotic expressions corresponding to the two limits $q^2/m^2\rightarrow 0$ and $m^2/q^2\rightarrow 0$ introduced above.  We also show the link to the Peskin and Takeuchi definition of $S$ in the Appendix~\ref{APPB}.

\subsection{Sending $q^2\rightarrow 0$ at fixed fermion masses}
We obtain for $q^2/m^2\rightarrow 0$
\be
\lim_{\frac{q^2}{m^2}\rightarrow 0} \delta S = \frac{17}{12} + \frac{317}{720 x} + \frac{919}{10080 x^2} + \mathcal{O}(x^{-3})\,\, ,\label{SPT}
\ee
where, as above, $x=\frac{m^2}{q^2}$.

\begin{figure*}[t]
\begin{center}
\subfigure[~$SU(3)$ with fundamental fermions.]{\includegraphics[width=\columnwidth]{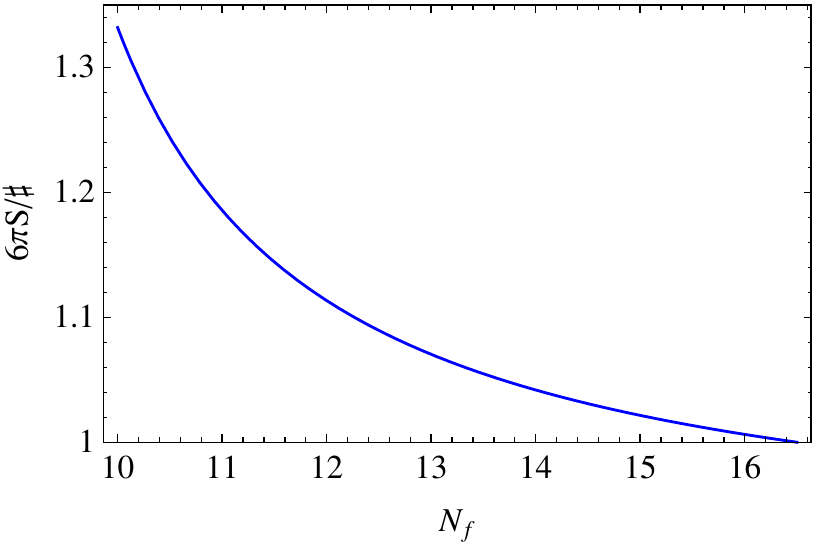}} \hfill
\subfigure[~$SU(2)$ with fundamental fermions.]{\includegraphics[width=\columnwidth]{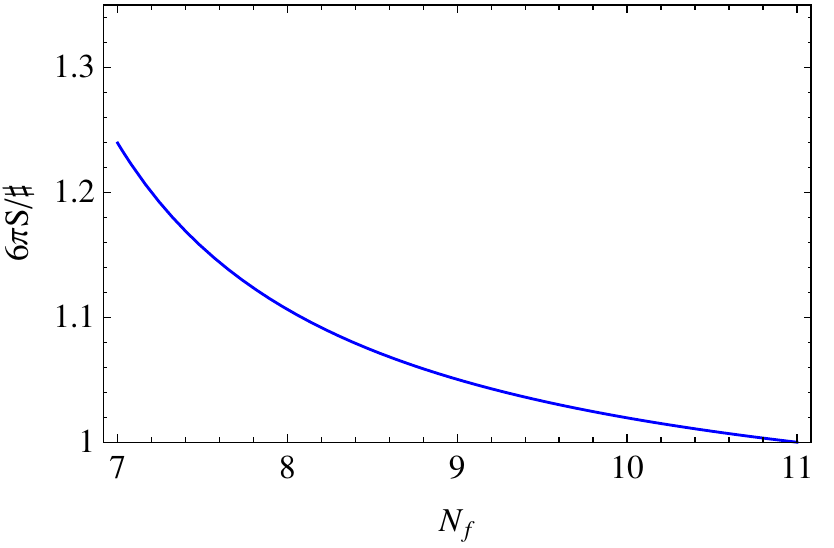}} \\
\subfigure[~$SU(3)$ with 2-index symmetric fermions.]{\includegraphics[width=\columnwidth]{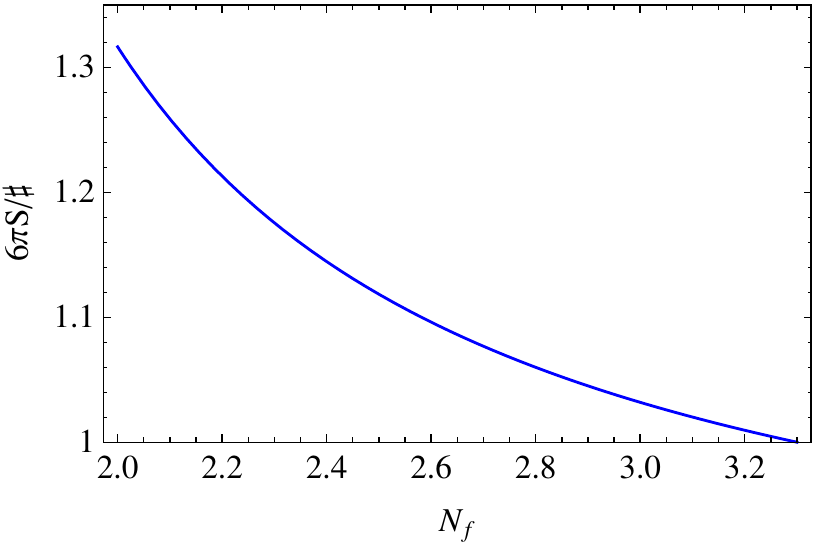}} \hfill
\subfigure[~$SU(2)$ with adjoint fermions.]{\includegraphics[width=\columnwidth]{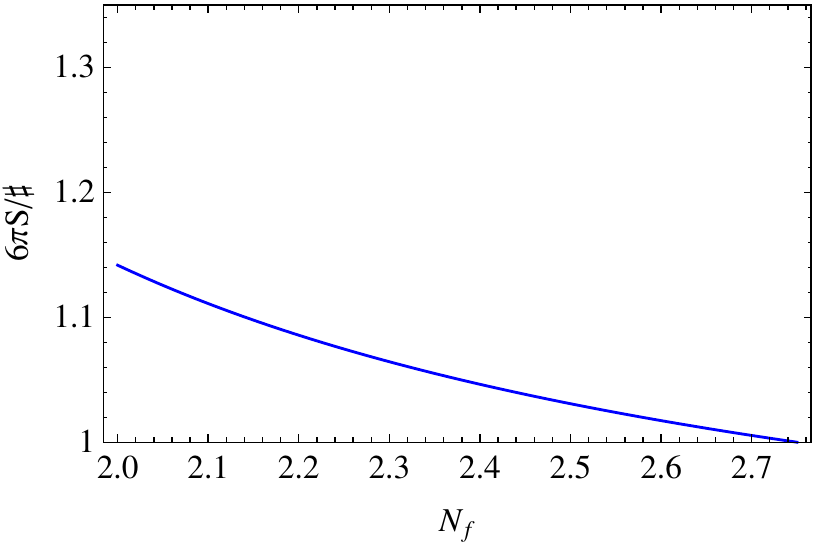}} 
\caption{Normalized conformal $S$-parameter near the perturbative upper bound of the conformal window for different theories.} 
\label{FIGSU3FUND}
\end{center}
\end{figure*}%

We evaluate $\alpha$ in \eqref{EQS2LOOP} at the energy corresponding to the common mass of the fermions taken to be much smaller than the technical scale $\Lambda_{U}$ above which the coupling constant stops walking and starts to run. For light fermions this is naturally the value of the coupling constant at the fixed point $\alpha^*$. It is perturbatively consistent to consider the 2-loops $\beta$-function to determine $\alpha$ at the fixed point. We have:
\be
\frac{\alpha^*}{4\pi}&=&-\frac{\beta_0}{\beta_1}\, ,\quad{\rm with}\\
\beta_0&=&\frac{11}3C_2\left[G\right]-\frac{4}3T\left[r\right] N_f\, , \\
\beta_1&=&\frac{34}3C_2^2\left[G\right]-\left( \frac{20}3 C_2\left[G\right]+4C_2\left[r\right]\right) T\left[r\right] N_f \, .
\ee 

Using this value for $\alpha$, the normalized $S$-parameter in the limit $q^2/m^2\rightarrow 0$ at 2-loops is then given by:
\be
\lim_{\frac{q^2}{m^2}\rightarrow 0} \frac{6\pi S}{\sharp } = 1 -\frac{17}{12}\frac{\beta_0}{\beta_1} C_2 \left[ r \right] \, , \label{EQSDEP}
\ee
where we kept only the leading order term in $1/x$. At this order, the $S$-parameter can also be re-expressed as a function of the 1-loop anomalous dimension of the mass $\gamma_m$ as
\be
\lim_{\frac{q^2}{m^2}\rightarrow 0} \frac{6\pi S}{\sharp } = 1 +\frac{17}{72}\gamma_m(\alpha^*)\, ,
\label{Sgamma}
\ee
with
\be
\gamma_m(\alpha)=\frac32 C_2\left[r\right] \frac{\alpha}{\pi}\, .
\ee
The above expressions show that the normalized $S$-parameter is a decreasing function of $N_f$ near the upper boundary of the conformal window. This result is in agreement with the conjecture formulated in \cite{Sannino:2010ca}. As an illustration we plot the normalized $S$-parameter, given in Eq.~\eqref{EQSDEP}, as a function of the number of fermions $N_f$ within the conformal window up to the critical number of fermions for which asymptotic freedom is lost in Fig.~\ref{FIGSU3FUND} for the cases of SU(3) with fundamental fermions and two-index symmetric fermions, and for SU(2) with fundamental and adjoint fermions.   

Note, however, that the unnormalized $S$ shows the opposite behavior, that is, it increases with the number of fermions. This statement holds in the perturbative regime and might happen that the full $S$ is not a monotonic function of the number of flavors. 

Clearly our estimate for the $S$-parameter is reliable only in the perturbative limit near the critical number of fermions above which asymptotic freedom is lost.

\subsection{ Taking $m^2\rightarrow 0$ first and the conformal limit}
In the opposite limit of $m^2/q^2\rightarrow 0$ we find:
\begin{eqnarray}
\lim_{\frac{m^2}{q^2}\rightarrow 0}\Re[\delta S] &=& -\frac{9 x}{4} \left[-7+2 \pi ^2+8 \zeta[3]-\right. \nonumber \\ 
&&\Bigl. - 2 \log (x) (3+\log (x)) \Bigr]\, , 
\end{eqnarray}
and
\be
\lim_{\frac{m^2}{q^2}\rightarrow 0}\Im[\delta S] = \frac{9\pi}{2} x \left(3+2 {\log}(x)\right)\, ,
\ee
for the real and imaginary part of $\delta S$ respectively.
This is consistent with the 1-loop result which shows that an imaginary part develops and correctly vanishes in the small mass limit at a finite value of $q^2$.

\begin{figure}[t]
\begin{center}
\includegraphics[width=\columnwidth]{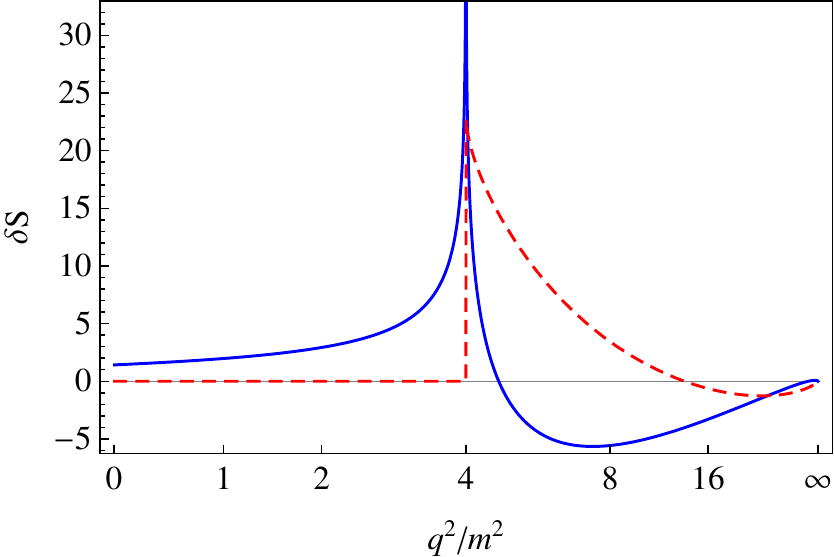} 
\caption{Real (blue, solid) and imaginary (red, dashed) parts of the 2-loop contribution $\delta S$ to the $S$-parameter as a function of $q^2/m^2$. To plot simultaneously the $q^2/m^2 \rightarrow 0$ and $\infty$ limits  we use a nonlinear scale for the horizontal axis which is proportional to $\arctan \left( q^2/m^2 \right) $.} 
\label{FIGlimit1}
\end{center}
\end{figure}

We then plot the complete 2-loops expressions for the real and imaginary parts of $\delta S$ in Fig.~\ref{FIGlimit1}. As for the one loop case the imaginary part of $S$ vanishes for $q^2/m^2<4$ while it is non zero above this kinematic threshold associated to particle creation.
At 2-loops a logarithmic divergence emerges in the real part at the same kinematic threshold.
The appearance of this logarithmic divergence in the perturbative expansion at order $\mathcal{O}(\alpha)$ is well known in the literature of QCD corrections to EW parameters, see e.g.~\cite{Chang:1981qq,Kniehl:1992dx}. The origin of this enhancement near the kinematic threshold of the 2-loop diagrams can be traced back to the Coulomb singularity~\cite{Kinoshita:1962ur}.

%
%

\section{On the $S$-parameter lower bound and the link to gauge duality}

As we decrease the number of flavors, within the conformal window, we have shown that
the normalized $S$ increases. This statement is under control in perturbation theory and lends further support to the claim made in \cite{Sannino:2010ca} according to which the unity value of the normalized $S$-parameter constitutes the absolute lower bound across the entire phase diagram. 

In formulae the $S$-parameter satisfies:
 \begin{equation}
S_{\rm norm} \equiv \frac{6\pi S}{\sharp} \geq 1  \quad {\rm when} \quad {\frac{q^2}{m^2} \rightarrow 0}  \ .
\label{sbound}
\end{equation}

Beyond perturbation theory it has also been shown \cite{Sannino:2010fh} that near the lower bound of the conformal window the $S$-parameter can be estimated via gauge duality~\cite{Sannino:2009qc,Sannino:2009me,Terning:1997xy}. There is, in fact, the fascinating possibility that generic asymptotically free gauge theories have magnetic duals. These are genuine gauge theories with typically a different gauge group with respect to the original electric theory and matter content. The full theory possesses, however, the same flavor symmetries. At low energy the electric and magnetic theory flow to the same infrared physics. 
The computation of the $S$-parameter would be then possible, in perturbation theory, near the lower bound of the conformal window since the dual gauge theory there is expected to be in a perturbative regime.

A candidate gauge theory dual to QCD in the conformal window, i.e.\ to an $SU(3)$ gauge theory with a large enough number ($N_f$) of Dirac flavors in the fundamental representation,  was put forward in \cite{Terning:1997xy,Sannino:2009qc,Sannino:2009me}.
The proposed dual theory is an $SU(X)$ gauge theory which possesses the same global symmetry group $SU_L(N_f)\times SU_R(N_f) \times U_V(1)$ of the original \textit{electric} theory. The elementary matter fields of dual theory are \textit{magnetic} quarks ${q}$ and $\widetilde{q}$ and gauge singlet Weyl fermions, which can be identified with the baryons of the electric theory. In addition, one is also free to add more elementary fields $M$ corresponding to the mesons of the electric theory, since they do not alter the global symmetries of the theory. Such fields are in fact needed to introduce interactions between the magnetic quarks and the gauge singlet fermions via Yukawa-type interactions.

\begin{table}[h]
\[ \begin{array}{|c| c|c c c|c|} \hline
{\rm Fields} &\left[ SU(X) \right] & SU_L(N_f) &SU_R(N_f) & U_V(1)& \# ~{\rm  of~copies} \\ \hline 
\hline 
 q &\Yfund &{\Yfund }&1&~~y &1 \\
\widetilde{q} & \overline{\Yfund}&1 &  \overline{\Yfund}& -y&1   \\
A &1&\Ythreea &1&~~~3& \ell_A \\
S &1&\Ythrees &1&~~~3& \ell_S \\
C &1&\Yadjoint &1&~~~3& \ell_C \\
B_A &1&\Yasymm &\Yfund &~~~3& \ell_{B_A} \\
B_S &1&\Ysymm &\Yfund &~~~3& \ell_{B_S} \\
{D}_A &1&{\Yfund} &{\Yasymm } &~~~3& \ell_{{D}_A} \\
{D}_S & 1&{\Yfund}  &{\Ysymm} &  ~~~3& \ell_{{D}_S} \\
\widetilde{A} &1&1&\overline{\Ythreea} &-3&\ell_{\widetilde{A}}\\
\widetilde{S} &1&1&\overline{\Ythrees} & -3& \ell_{\widetilde{S}} \\
\widetilde{C} &1&1&\overline{\Yadjoint} &-3& \ell_{\widetilde{C}} \\
M^i_{j} &1&\Yfund &\overline{\Yfund} & 0 &1 \\
 \hline \end{array} 
\]
\caption{Massless spectrum of {\it magnetic} quarks and baryons and their  transformation properties under the global symmetry group. The last column represents the multiplicity of each state and each state is a  Weyl fermion.}
\label{dualgeneric}
\end{table}

The spectrum of the proposed magnetic dual of QCD is summarized in Table~\ref{dualgeneric}.
The multiplicity of each baryonic field in the spectrum is denoted by the $\ell$s.
The parity and charge conjugation symmetry of the underlying theory requires $\ell_{J} = \ell_{\widetilde{J}}$~ with $J=A,S,C$ and $\ell_B = - \ell_D$. 

 Near the lower end of the conformal window the {\it magnetic} S-parameter, i.e. the S-parameter computed in the magnetic theory, is \cite{Sannino:2010fh}: 
 \begin{eqnarray}
S_{m} & = &S_q + S_B +S_{\rm M} \ ,
\end{eqnarray}
with $S_q$, $S_B$ and $S_{\rm M}$ the contributions coming from the magnetic quarks, the baryons and the mesons respectively. 
 In \cite{Sannino:2010fh} it was considered the case in which we gauge, with respect to the EW interactions, only the $SU_L (2)\times SU_R (2)$ subgroup where the hypercharge is the diagonal generator of $SU(2)_R$.  In this case only one doublet contributes directly to the $S$-parameter and we have  \cite{Sannino:2010fh}: 
\begin{eqnarray}
\frac{6\pi}{3}S_{m} = \frac{X}{3} + \frac{\ell_C + \ell_{B_A}}{3} + \frac{25}{729} \, \ell_{B_S} \left( 32 \log2 - 39 \right) - 0.14 \ . \nonumber \\
\end{eqnarray} 

Using this expression for a possible QCD dual provided in \cite{Sannino:2009qc} for which $X = 2 N_f - 15$, $\ell_A = 2$, $\ell_{B_A} = -2$ with the other $\ell$s vanishing, we can obtain an estimate of the $S$-parameter. Asymptotic freedom of the dual theory requires $N_f\ge9$. For $N_f=9$ we obtain $6\pi S_{m}/3 = 1.523$, while for $N_f = 10$ we have $6\pi S_{m} /3= 2.19$.
It is quite remarkable that the computation in the magnetic theory in \cite{Sannino:2010fh} yields an estimate which is consistent with the lower bound and the perturbative computations presented here. 

{\it How can we connect the conformal $S$ with the one below the conformal window?} 

As we decrease the number of flavors we cross into the chirally broken phase and conformality is lost. Below the critical number of flavors corresponding to the lower bound of the conformal window, a dynamical mass of the fermions is generated. In the broken phase we should compute the $S$-parameter, in the zero momentum limit, with the hard mass of the fermions replaced by the hard plus the dynamical one. We noted in \cite{Sannino:2010ca} that this indicates that the broken and symmetric phases are smoothly connected when discussing the normalized $S$-parameter.

\begin{figure}[t]
\begin{center}
\includegraphics[width=\columnwidth]{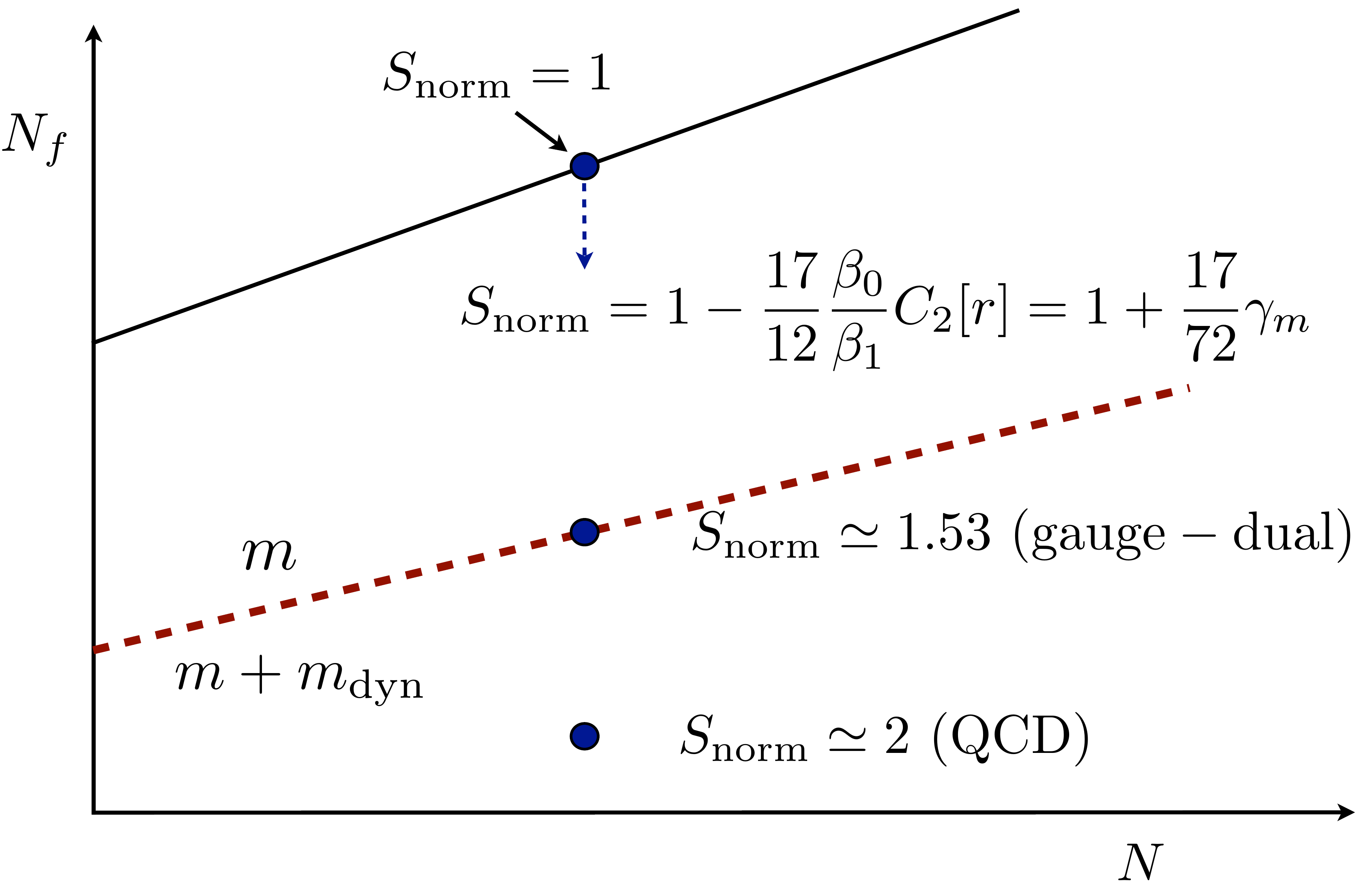} 
\caption{Cartoon of the dependence of the normalized $S$-parameter ($S_{\rm norm}$) on the number of Dirac flavors transforming according to the fundamental representation of the $SU(3)$ gauge theory across the phase diagram. The solid oblique line corresponds to the points where the theory looses asymptotic freedom. Chiral symmetry breaks below the dashed line while the conformal window is between the two lines. $S_{\rm norm} =1$ at the upper end of the conformal window and it increases according to the formulae  \eqref{EQSDEP} and \eqref{Sgamma}  when decreasing the number of flavors. This result is trustable within the perturbative regime. The estimate at the lower end of the conformal window has been derived using gauge duality in \cite{Sannino:2010fh}.    The QCD value is reported too. Below the conformal window a dynamical mass $m_{\rm dyn}$ is generated (on the top of the bare mass $m$) and it is expected to vanish smoothly across the lower boundary suggesting that the $S$-parameter is smooth too.} 
\label{S-cartoon}
\end{center}
\end{figure}

Therefore we expect the lower bound on the normalized $S$-parameter to apply to the entire phase diagram concerning asymptotically free gauge theories.  We elucidate the above picture in Fig.~\ref{S-cartoon}.


Note that a lower bound on the normalized $S$-parameter is compatible with the previous claims \cite{Appelquist:1998xf, Kurachi:2006mu,Kurachi:2006ej,Kurachi:2007at} that in a near conformal theory the value of $S$ can be smaller than the one obtained in QCD. However from our results we do not expect a negative $S$-parameter to occur in an asymptotically free gauge theory. While we work in a controlled regime in which our prediction for the flavor dependence of the $S$-parameter is trustable we note that such a dependence has been long sought after. In fact many estimates have been provided in the literature using various approximations in field theory  \cite{Sundrum:1991rf}  or using computations inspired by the original AdS/CFT correspondence \cite{Maldacena:1997re} in \cite{Hong:2006si,Hirn:2006nt,Piai:2006vz,Agashe:2007mc,Carone:2007md,Hirayama:2007hz}. Recent attempts to use AdS/CFT inspired methods can be found in \cite{Dietrich:2009af,Dietrich:2008up,Dietrich:2008ni,Hirn:2008tc,Nunez:2008wi,Fabbrichesi:2008ga,Anguelova:2010qh}.

The results obtained in the limit of sending to zero the mass of the fermions at a nonzero external momentum is also interesting since it applies immediately to models of unparticle physics with unparticle matter gauged under the weak interactions.  


{ 
The relevance of this section has been, so far, to demonstrate consistency of the behavior of the normalized $S$-parameter flavor dependence near the upper and the lower end of the conformal window. 

It is natural to ask what happens if we start increasing the number of flavors near  the lower end of the conformal window. Using the dual one can still, in principle, use a perturbative expansion in the magnetic coupling at the infrared fixed point to investigate the $S$-parameter behavior in this region. This computation is, however, considerably more involved than the one performed for the electric theory and will be presented elsewhere. For the scope of this initial work we simply illustrate the feasibility of this computation and some of its salient aspects by providing a specific two-loops contribution which can be immediately determined from our work in the earlier section. More precisely we compute, for the first time, the contribution of the magnetic fermions, to the two-loops order in the magnetic theory, for the $S$-parameter which reads: 
\begin{equation}
\frac{6\pi}{3}S_q = \frac{X}{3} \left[ 1 + \frac{17}{12}\frac{\alpha^{\ast}_m}{4\pi} \frac{X^2 - 1}{2X}\right] \ .
\end{equation}  
with $\alpha_m^{\ast}={g^{\ast}}^2_m/ 4\pi$ the value of the {\it magnetic} coupling constant at the infrared fixed point. At the fixed point the famous Dirac inspired electric-magnetic duality requires: 
\begin{equation}
g^{\ast} \cdot \, g^{\ast}_m \sim {\rm constant} \ .
\end{equation}
and therefore 
\begin{equation}
\alpha_m^{\ast} \sim \frac{1}{\alpha} \ .
\end{equation}
The beauty of this relation is that clearly shows how small the perturbative corrections to $S_q$ become near the lower end of the conformal window because of the fact that the electric theory coupling constant becomes large. Although this is only a small fraction of the full two-loops corrections to the normalized magnetic $S$-parameter it suggests the emergence of an intriguing  picture in which the electric-magnetic gauge duality idea shows its potential large impact in the understanding of nonperturbative gauge dynamics. 
}

Our present results further strengthen the lower bound conjecture \cite{Sannino:2010ca} and therefore favor, from the precision EW constraints point of view, technicolor models with the smallest number of techniflavors  gauged under the EW symmetry  \cite{Sannino:2004qp, Hong:2004td,Dietrich:2005jn,Dietrich:2005wk,Evans:2005pu,Christensen:2005cb,Gudnason:2006ug,Gudnason:2006yj,Foadi:2007ue,Foadi:2008xj,Belyaev:2008yj,Foadi:2008ci,Foadi:2007se}. These include models of partially gauged technicolor \cite{Dietrich:2005jn,Dietrich:2006cm,Fukano:2009zm,Frandsen:2009fs} in which only two techniflavors are EW gauged. 

We can straightforwardly extend the present findings to the case in which different matter representations are considered. An example is ultra minimal walking technicolor \cite{Ryttov:2008xe}. In fact, the effects of the fermion transforming according to the matter representation, which is  singlet with respect to the SM interactions,  at the two-loops level affects only the value of coupling at the IRFP while the functional form of the normalized $S$-parameter \eqref{EQSDEP} remains unchanged. The presence of the extra matter representation is to push the IRFP closer to the perturbative regime  thereby reducing, for a given number of flavors gauged under the EW, the associated $S$-parameter. Needless to say  the universal bound still holds. The generalization to symplectic and orthogonal technicolor gauge groups \cite{Sannino:2009aw} is straightforward and the results interesting since {\it orthogonal} technicolor models \cite{Frandsen:2009mi} have already been proposed in the literature. 

In the future we plan to generalize the present analysis at nonzero temperature, matter density, and finite volume. 
\section{Conclusions}

 The 2-loop results presented here provide a natural benchmark for lattice computations~\cite{Catterall:2007yx,Appelquist:2007hu,DelDebbio:2008wb,Shamir:2008pb,Deuzeman:2008sc,DelDebbio:2008zf,Catterall:2008qk,Svetitsky:2008bw,DeGrand:2008dh,Fodor:2008hm,Fodor:2008hn,Deuzeman:2008pf,Deuzeman:2008da,Hietanen:2008vc,Jin:2008rc,DelDebbio:2008tv,DeGrand:2008kx,Fleming:2008gy,Hietanen:2008mr,Appelquist:2009ty,Hietanen:2009az,Deuzeman:2009mh,Fodor:2009nh,DeGrand:2009mt,DeGrand:2009et,Hasenfratz:2009ea,DelDebbio:2009fd,Fodor:2009wk,Fodor:2009ar,Appelquist:2009ka,DeGrand:2009hu,Catterall:2009sb,Bursa:2009we,Lucini:2009an,Pallante:2009hu,Bilgici:2009kh,Machtey:2009wu,Moraitis:2009xt,Kogut:2010cz,Hasenfratz:2010fi,DelDebbio:2010hu,DelDebbio:2010hx} of the $S$-parameter for vector-like gauge theories featuring an IRFP.  
To be specific  we suggest to study the $S$-parameter for $SU(3)$ gauge theory with 16  and 12 fundamental flavors  on the lattice and to compare the results with our perturbative predictions. This comparison will serve as a relevant test of the hypothesis of conformality in a controllable manner. Deviations from the perturbative estimate and the absolute lower bound \cite{Sannino:2010ca} can be tested for any gauge theory investigated on the lattice such as the phenomenologically relevant (Next) Minimal Walking Technicolor \cite{Sannino:2004qp,Dietrich:2005jn} models.  

Furthermore by determining the value of the $S$-parameter on the lattice one can test weak-strong gauge duality as suggested in \cite{Sannino:2010fh}.

Our results lend support to the existence of a universal lower bound for the normalized $S$-parameter \cite{Sannino:2010ca} which can be used to identify  models of dynamical EW symmetry breaking and unparticle physics not in contradiction with EW precision measurements.

\acknowledgments
C.P. and F.S. thank the CERN Theory Institute for its kind hospitality during the meeting "Future directions in lattice gauge theory - LGT10" where this work has been finalized. { While this Letter was under review two of the authors of this Letter together with Mojaza and Nardecchia in \cite{Mojaza:2011rw} have provided further evidence for the existence of gauge duals in nonsupersymmetric gauge theories such as QCD with one adjoint fermion. The dual passes a remarkably large number of consistency conditions and it is valid for {\it any} number of colors within the conformal window. }
\\ \\
\appendix
\section{2-loops expression for the $S$-parameter\label{APPA}}
In this Appendix we give the complete expression for the 2-loops contribution to the $S$-parameter defined in Eq.~\eqref{eq:Sdefinition}. The formula for $\delta S$  given in Eq.~\eqref{EQS2LOOP} has been obtained using the results of the QCD corrections to the EW gauge boson self-energies given by Djouadi and Gambino \cite{Djouadi:1993ss}. {}For equal up- and down-type fermions masses $M_1=M_2=m$ , the expression for $\delta S$ reads:
\begin{widetext}
\begin{eqnarray}
\delta S&=&\frac{3 x}{4}\left[12 (2 x-1) \left(\text{Li}_3\left(y^2\right)+4 \text{Li}_3(y)+2 \zeta (3)\right)-8 \sqrt{1-4 x} \left(\text{Li}_2\left(y^2\right)+2 \text{Li}_2(y)\right)-4 x+21\right.\nonumber\\
&+&2 \log (-y) \left((8-16 x) \left(\text{Li}_2\left(y^2\right)+2 \text{Li}_2(y)\right)-\sqrt{1-4 x} (8 \log (1-y)+16 \log (1+y)+2 x-9)\right)\nonumber\\
&+&\left.2 \log ^2(-y) \left((4-8 x) \left(2 \log \left(1-y^2\right)-\log (1-y)\right)+2 x (x+2)+6 \sqrt{1-4 x}-3\right)\right]\,\, ,
\end{eqnarray}
\end{widetext}
where
\beq
y=\frac{4 x}{\left(\sqrt{1-4 x}+1\right)^2}\,,\ x=\frac{m^2}{q^2}\,\, ,
\eeq
$q$ is the external momentum flowing in the vacuum polarization diagrams and $\text{Li}_n(z)=\sum_{k=1}^{\infty}z^k/k^n$ is the polylogarithm function. We stress, however, that these results were specialized for the perturbative computations of the S-parameter in the conformal window for the first time here. 

\section{Peskin - Takeuchi $S$-parameter\label{APPB}}
The $S$-parameter as defined by Peskin and Takeuchi (PT) in \cite{Peskin:1990zt}
\be S_{\rm PT}&=&-16\pi\frac{\partial\Pi_{3Y}(q^2)}{\partial q^2}  \bigg|_{q^2=0} \,,
\ee
can be easily recovered from the one defined by He, Polonsky and Su~\cite{He:2001tp}, in the limit $q^2\rightarrow 0$ of \eqref{eq:Sdefinition}.
Explicitly, at one loop from \eqref{eq:Sfermion} we have:
\be
S_{\rm PT} = \frac{\sharp}{6\pi}\left\{1-4Y\log\left(\frac{M_1}{M_2}\right)\right\} \, .
\ee
At 2-loops the expression for $S_{\rm PT}$ is easily obtained from the \eqref{EQS2LOOP} and \eqref{SPT} for the special case of degenerate fermion masses $M_1=M_2$, while in the general non-degenerate case we obtain:
\be
\delta S_{\rm PT}=\frac{17}{12}-3Y\log\left(\frac{M_1}{M_2}\right)\, .
\ee

\end{document}